\documentclass[12pt,preprint]{aastex}
\newcommand{\kms}{km~s$^{-1}$}
\newcommand{\ergss}{ergs~s$^{-1}$}
\newcommand{\lxlo}{$L_X / L_{opt}$}
\newcommand{\chandra}{{\sl Chandra }}
\newcommand{\OII}{[O{\scshape ii}] }
\newcommand{\NII}{[N{\scshape ii}] }

\begin{document}
				\title{
\chandra Observations of Point Sources in  Abell~2255
				}

				\author{
David S. Davis\altaffilmark{1,}\altaffilmark{2}}

				\author{
Neal A. Miller\altaffilmark{3}}

\and
				\author{
Richard F. Mushotzky\altaffilmark{4}    
				}
\altaffiltext{1}{Laboratory for High Energy Astrophysics, Code 661,
Greenbelt, MD 20771}
\altaffiltext{2}{Joint Center for Astrophysics, Department of Physics, 
University of Maryland,
Baltimore County, 1000 Hilltop Circle, Baltimore, MD 21250 }
\altaffiltext{3}{NASA/GSFC UV/Optical Branch, Code 681,
Greenbelt, MD 20771}
\altaffiltext{4}{Laboratory for High Energy Astrophysics, Code 662,
Greenbelt, MD 20771}

%				\clearpage

                                \begin{abstract}
In our search for ``hidden'' AGN 
we present results from a \chandra observation of the nearby cluster
Abell~2255. Eight cluster galaxies are associated with point-like
X-ray emission, and we classify these galaxies based on their X-ray,
radio, and optical properties. At least three are associated with
active galactic nuclei (AGN) with no optical signatures of nuclear
activity, with a further two being potential AGN. Of the potential
AGN, one corresponds to a galaxy with a post-starburst optical
spectrum. The remaining three X-ray detected cluster galaxies consist
of two starbursts and an elliptical with luminous hot gas. Of the
eight cluster galaxies five are associated with luminous (massive)
galaxies and the remaining three lie in much lower luminosity
systems. We note that the use of X-ray to optical flux ratios for
classification of X-ray sources is often misleading, and strengthen
the claim that the fraction of cluster galaxies hosting an AGN based
on optical data is signficantly lower than the fraction based on X-ray
and radio data.

				\end{abstract}

				\clearpage

				\section{
Introduction
				}

Until fairly recently it has been thought that, at least at low
redshift, the relative fraction of active galaxies in rich clusters
was considerably lower than in the field \citep{dg83,yee93,dressler99}.  
Recent \chandra observations have revealed a more substantial
population of active galactic nuclei (AGN) in clusters of galaxies
\citep{cappi01,sun02,molnar02,martini02}.  Optically, the hosts of
these sources often appear relatively unremarkable: they have colors
consistent with cluster ellipticals, and optical spectra lacking
emission lines. Thus, traditional optical studies overlook these
sources and arrive at a fraction of $\sim1\%$ of all cluster galaxies
harboring an AGN \citep[e.g.,][]{dressler99}. Including these X-ray
selected AGN, however, yields a fraction of $\sim5\%$, consistent with
the fraction of field galaxies hosting an AGN on the basis of optical
studies \citep{barger02}.

Interestingly enough, similar results have been found via radio
continuum studies. The classical radio galaxies, with large jets and
lobes of radio emission, are one such example. Yet these sources are
rare, and typical clusters of galaxies only contain up to a few such
radio galaxies \citep{ledlow95}. However, more sensitive observations
have continued to reveal fainter compact radio sources in elliptical
galaxies, most of which are thought to be AGN \citep[e.g.,][]{sadler89}, 
and most of which have optical spectra
exhibiting no emission lines or weak \NII emission \citep{phillips86}.
The incidence of such AGN in clusters of galaxies appears to be consistent 
with that inferred from the recent X-ray studies \citep{miller02}.

The importance of these AGN is potentially significant.
\citet{martini02} note that a large population of cluster AGN would do
much to explain the hard X-ray background. Variations in the X-ray
background across the sky \citep[e.g.,][]{cowie02} would be a natural
consequence of cosmological structure and the cluster
distribution. Furthermore, studies based on optical properties of
galaxies are clearly overlooking a large number of active
galaxies. This complicates analyses of evolutionary trends. Evidence
for this is seen in recent X-ray surveys \citep{barger02,hasinger03}
which show that X-ray selected objects evolve rather differently from
optically selected ones and that the space density of X-ray selected
AGN is considerably larger than that of optically selected objects.

Clusters of galaxies are attractive for studying galaxy evolution
because they contain large numbers of galaxies all at the same
redshift. Thus, samples of AGN derived from a given cluster are
subject to the same selection effects. Comparisons among the
cluster-selected AGN may thereby be used to investigate questions such
as the nature of the X-ray selected, optically ``dull'' galaxies
\citep{elvis80}. In addition, we seek to address if and why the
fraction of optically-selected AGN in clusters differs from that in
the field. As data for more clusters becomes available, the question
of why the environment of optically-selected objects changes rapidly
with redshift \citep{yee93} may be addressed.

The question of how the cluster enviroment affects the nuclear 
activity is also an interesting question. 
Nuclear activity in galaxies is thought to be powered by accretion
onto a supermassive black hole. Over the long term this accretion
must be fueled by gas on the galactic scale transported into the 
inner few kpc of the galaxy. This transport can be triggered by 
mergers and interactions of galaxies \citep[e.g.,][]{hernquist89}.
The interactions of major clusters may also increase the activity
of member galaxies \citep{mru00}. 

Abell~2255 is a rich cluster which shows several signs that it is
undergoing a merger event. Some of the first evidence for a merger
came from early $Einstein$ X-ray data: \citet{jones84} found that
Abell~2255 has one of the largest X-ray core radii in their cluster
sample. \citet{stewart84} further analyzed $Einstein$ IPC data and
concluded that Abell~2255 does not contain a cooling flow, which is
unusual for such a rich cluster \citep{edge92}. Under certain
conditions, major mergers may disrupt pre-existing cooling flows and
increase the core radius of X-ray gas \citep{gomez02}. More recently,
analyses of $ROSAT$ data have confirmed the X-ray elongation and noted
that the X-ray peak of Abell~2255 is not centered on the brightest
cluster galaxies but is offset by $\sim$2\arcmin{}
\citep{burns95,feretti97}. Thus, the X-ray data indicate that this
cluster has recently undergone or is currently undergoing a
merger. Optically, Abell~2255 has a very large velocity dispersion,
$\approx$ 1200 \kms, and shows evidence for kinematical substructure in the
form of several associated groups \citep{hill03}. The large ratio of
velocity dispersion to X-ray temperature \citep[6.3 keV,][]{horner01}
also indicates a non-relaxed system. Furthermore, Abell~2255 contains
two comparably bright central dominant galaxies, which is reminiscent
of the Coma cluster and indicative of a merger of separate clusters
\citep{davis93,bird94}. Abell~2255 is also one of the rare clusters
which contains a radio halo \citep{jaffe79,hanisch82,feretti97}. To
date, radio halos have only been detected in clusters which exhibit
signatures of mergers.

Here we present \chandra data for the relatively nearby cluster
Abell~2255 ($z=0.08$) showing that eight of the galaxies in the central
region of the cluster are unresolved or small X-ray sources. This
cluster represents an excellent opportunity to revisit the question of
X-ray selected AGN in clusters, as a deep radio continuum survey of
the cluster has recently been performed \citep{miller03} and high
quality optical spectra are available through the Sloan Digital Sky
Survey Early Data Release \citep[SDSS,][]{stoughton02}. Thus, the
active galaxies in an X-ray selected sample may be evaluated over a
broad wavelength range in order to better assess the origin of the
activity and how it might be surmised.

				\section{
Data and Analysis              
				}

                                \subsection{
X-Ray
                                }

Abell 2255 was observed with the ACIS-I detector \citep{garmire03} on
the \chandra observatory for a total of 39 ksec on October 20 - 21
2000. Examining the lightcurve from chip 7 (the back illuminated CCD)
we excluded a small background enhancement at the beginning of the
exposure. This left a total of $\sim$34.7 ksec of exposure time.

\begin{figure}
\plotone{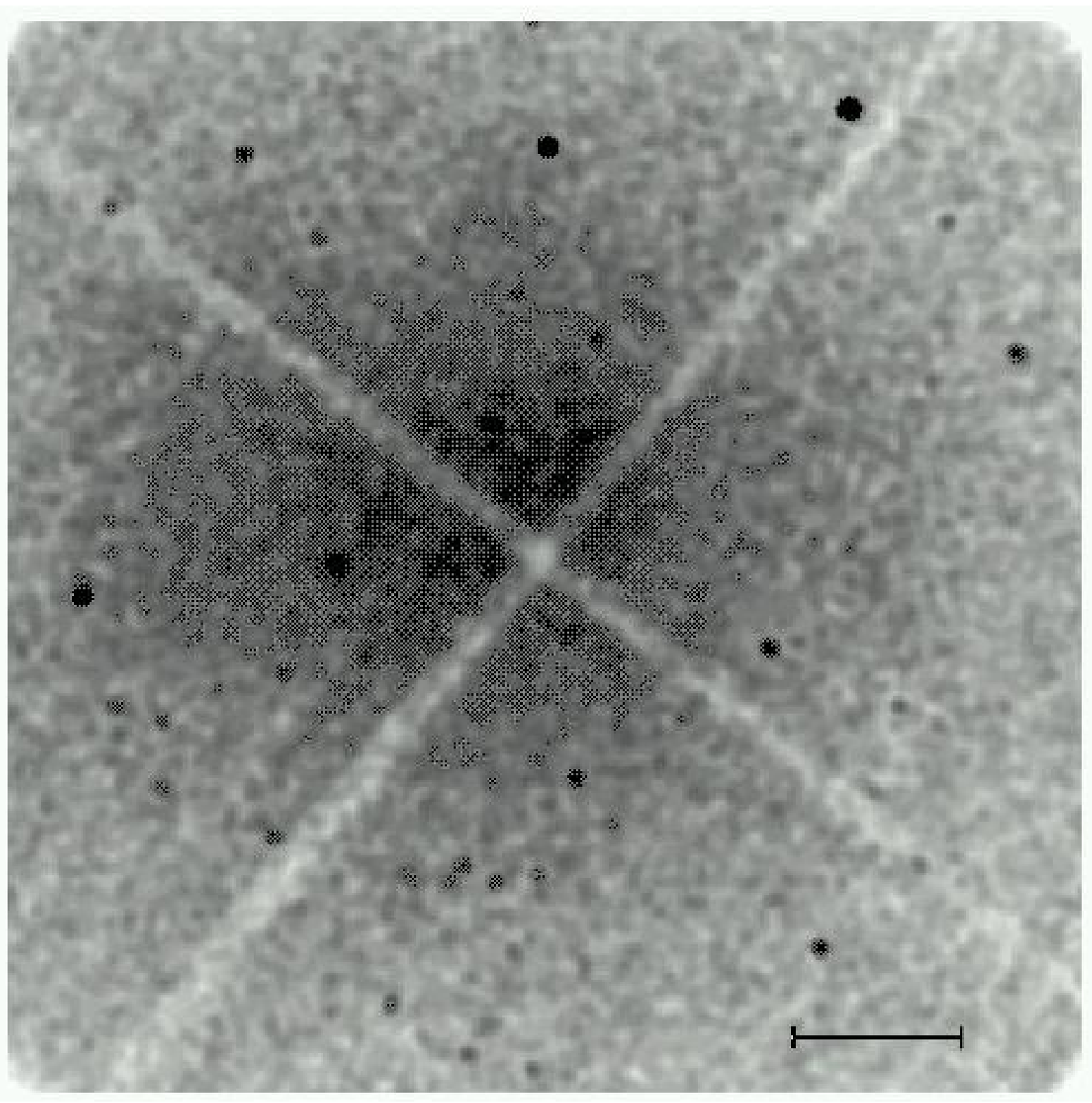}
%\plotone{a2255_xray_sm.eps}
\caption{The exposure corrected and smoothed X-ray image of the
Abell~2255 cluster in the 0.3 -- 10 keV band. The bar in the lower
right corner is 2\arcmin{} long and corresponds to 0.25 Mpc at the
distance of the cluster. The image is oriented with north up and east
to the left.  The image is binned by a factor of 4 (1 pixel is
2\arcsec{}) for display.  The smoothing and constrast is choosen to
emphasize point sources.
\label{sm_xray}\label{fig-xraysmooth} }
\end{figure}

\begin{figure}
%\plotone{a2255_xray_radio_cont.ps}
\plotone{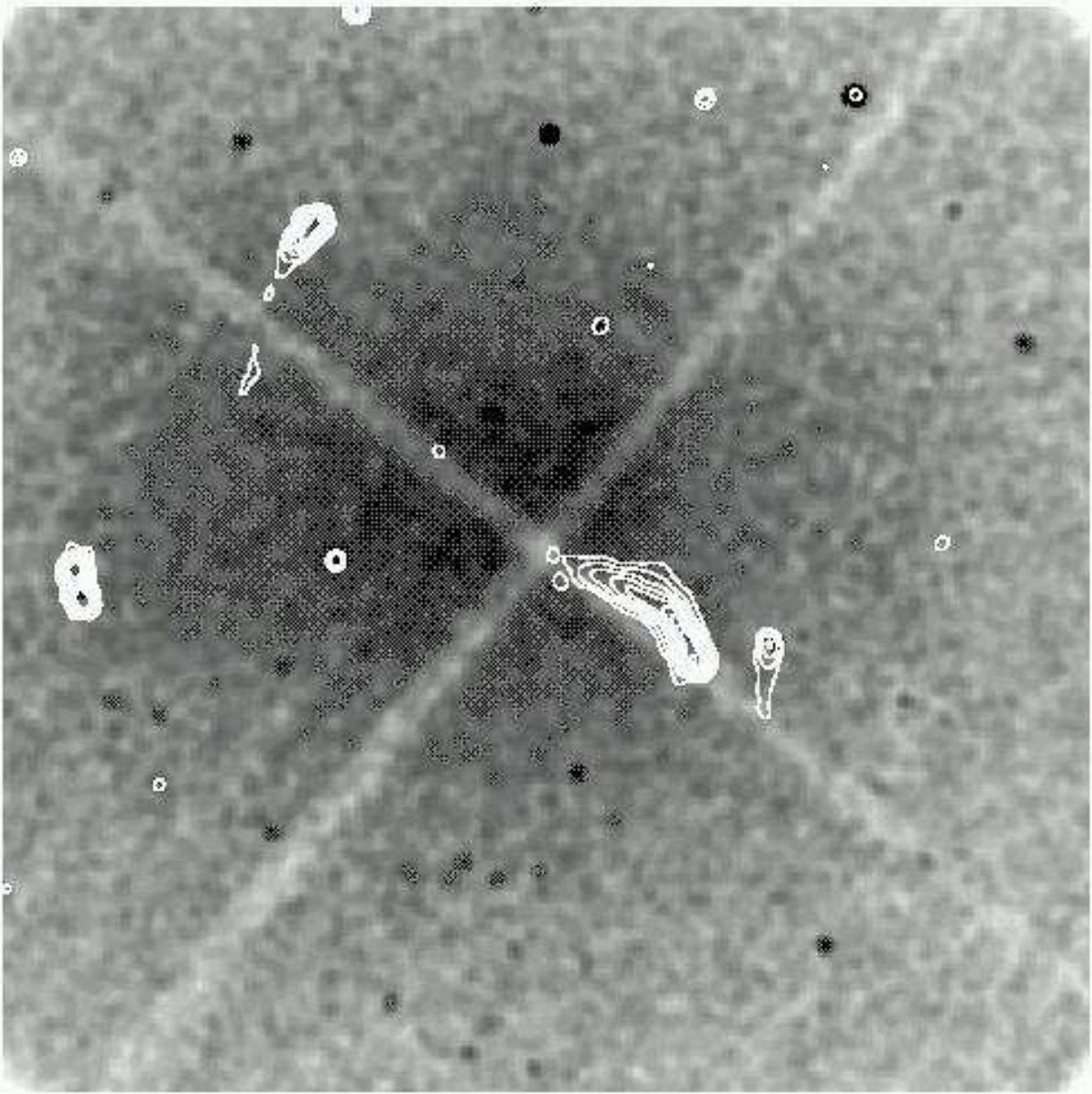}
\caption{
The smoothed X-ray image of Abell~2255 in the 0.3 to 
10.0 keV band. The data for each of the chips is corrected 
for the exposure time for each chip. The radio contours are from
\citet{miller03}.\label{fig-xrayrad}
}
\end{figure}

X-ray images of the Abell 2255 field are shown in Figures
\ref{fig-xraysmooth} and \ref{fig-xrayrad}. In the first figure, the
data have been smoothed with a Gausssian with a $\sigma$ of 1\arcsec{}
to emphasize the point sources, which may be seen along with the
diffuse X-ray emission from the cluster. The second figure depicts the
X-ray emission along with the radio contours \citep[][see Section
\ref{sec-radio}]{miller03}. We used the CIAO program WAVDETECT to
detect X-ray point sources in the field and visually confirmed the
detections. We used a sensitivity threshold of 10$^{-6}$, which gives 
about 1 false detection per 1024x1024 ACIS chip if
background is spatially uniform (Freeman et al. 2002).
These sources were checked against the SDSS data and the
velocity data from \citet{hill03} for the region. Only sources that
lay in the velocity range of 20,289 -- 27,705 \kms{} were identified
as cluster sources \citep[i.e., sources within $\pm 3\sigma$ of the
cluster systemic velocity, using the biweight estimators of
][]{beers90}.  This resulted in a list of eight galaxies in the
cluster center, which are listed in Table \ref{tbl-1} and henceforth
referred to by the ID numbers designated in that table. The columns in
table~\ref{tbl-1} are the ID, the source name, the power law index,
the X-ray luminosity in the 0.5 -- 10 keV band (assuming H$_0$ = 50 km/s/Mpc,
q$_0$=0.), the 20cm radio flux, the r$^*$ magnitudes and g$^*-$r$^*$ colors 
are determined from the SDSS data. The radial velocities we used are in the 
final column. 

We used {\sl XSPEC~11.2.o} software \citep{arnaud96} to fit a spectral
model to the extracted spectra.  We extracted the spectra using an
aperture with a radius of 10\arcsec{} for the point sources and
20\arcsec{} for the extended source to obtain all the flux. The
background for each spectrum was obtained from a ring from outside the
source region. This removes the underlying cluster emission as well as
the normal background components. Finally, the extracted spectra were
rebinned so that each channel had a minimum of 25 counts.

For the brightest source (\#1 in Table \ref{tbl-1}) we fit the
spectrum with a power law plus a contribution from a soft
thermal component associated with the galaxy. All fits included a
component for the extra absoption on the ACIS detectors (acisabs in
XSPEC) and a variable absorption component due to the column density
of Galactic hydrogen in the line-of-sight. The redshift of the model
spectrum was fixed to the heliocentric velocity of the cluster.  For
the other sources with sufficient counts we fit a simple power law
model, or if there were too few counts for spectral fitting we
estimated the flux by fixing the index of the power law model to
1.7. The extracted spectra were fit between $\sim$0.5 and 9.0 keV,
with the exact energy boundaries being set by the channel
grouping. Once a minimum in $\chi^2$ was found, the 90\% confidence
errors were determined for the free parameters. For the weaker
sources, hardness ratios were calculated in the 0.5--1.5, 1.5--2.5,
and 2.5--4.0 keV bands and compared to simple thermal and power law
models.

\begin{figure}
%\plotone{src1_xray_spec.ps}
\plotone{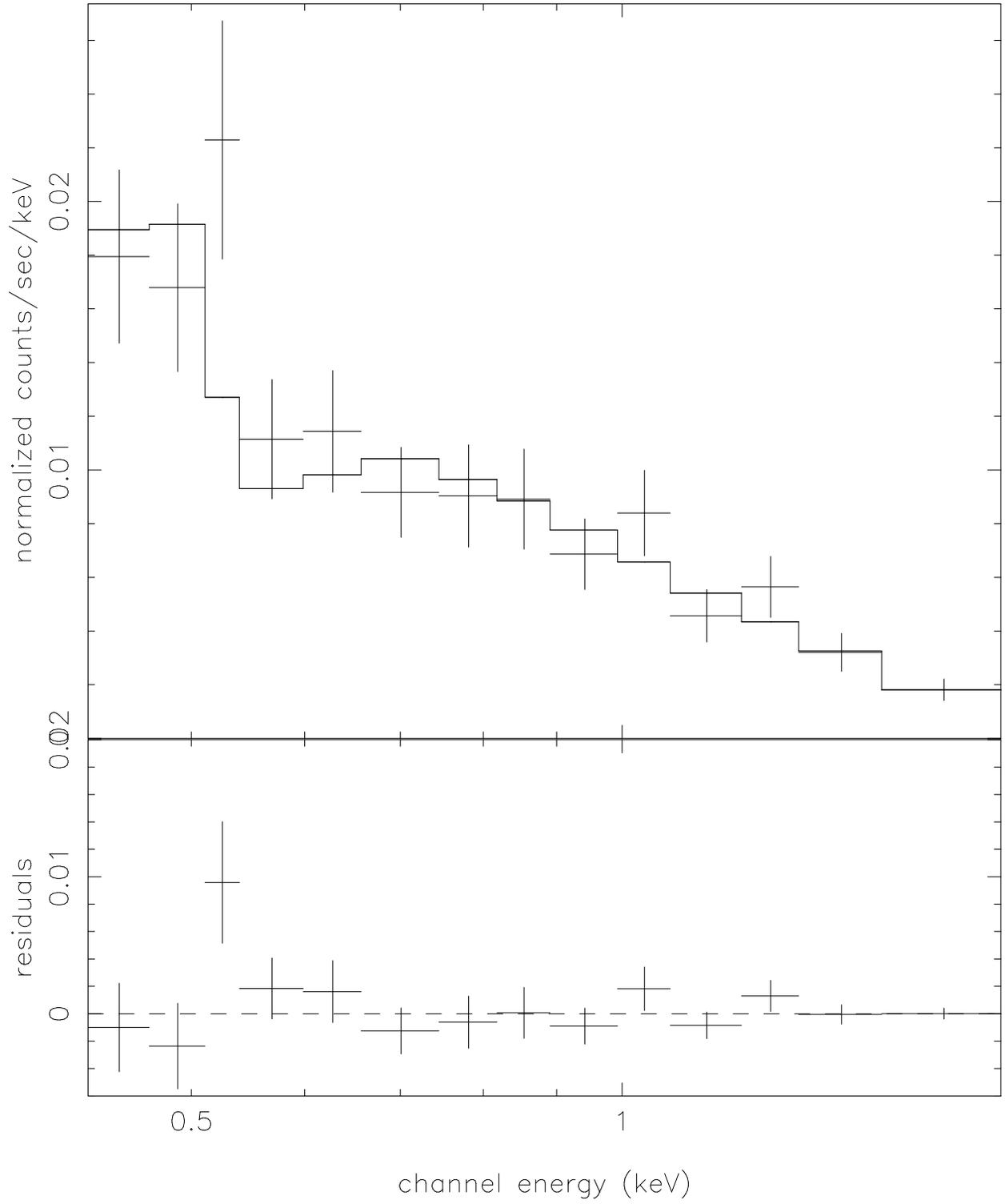}
\caption{The extracted X-ray spectrum for source \#1. The crosses are the 
data points and the solid line is the best fit power law plus a thermal
plasma model. \label{n1_xray}
}
\end{figure}

\begin{figure}
%\plotone{a2255_src1_KPNO_xray_cmky.ps}
\epsscale{0.65}
\plotone{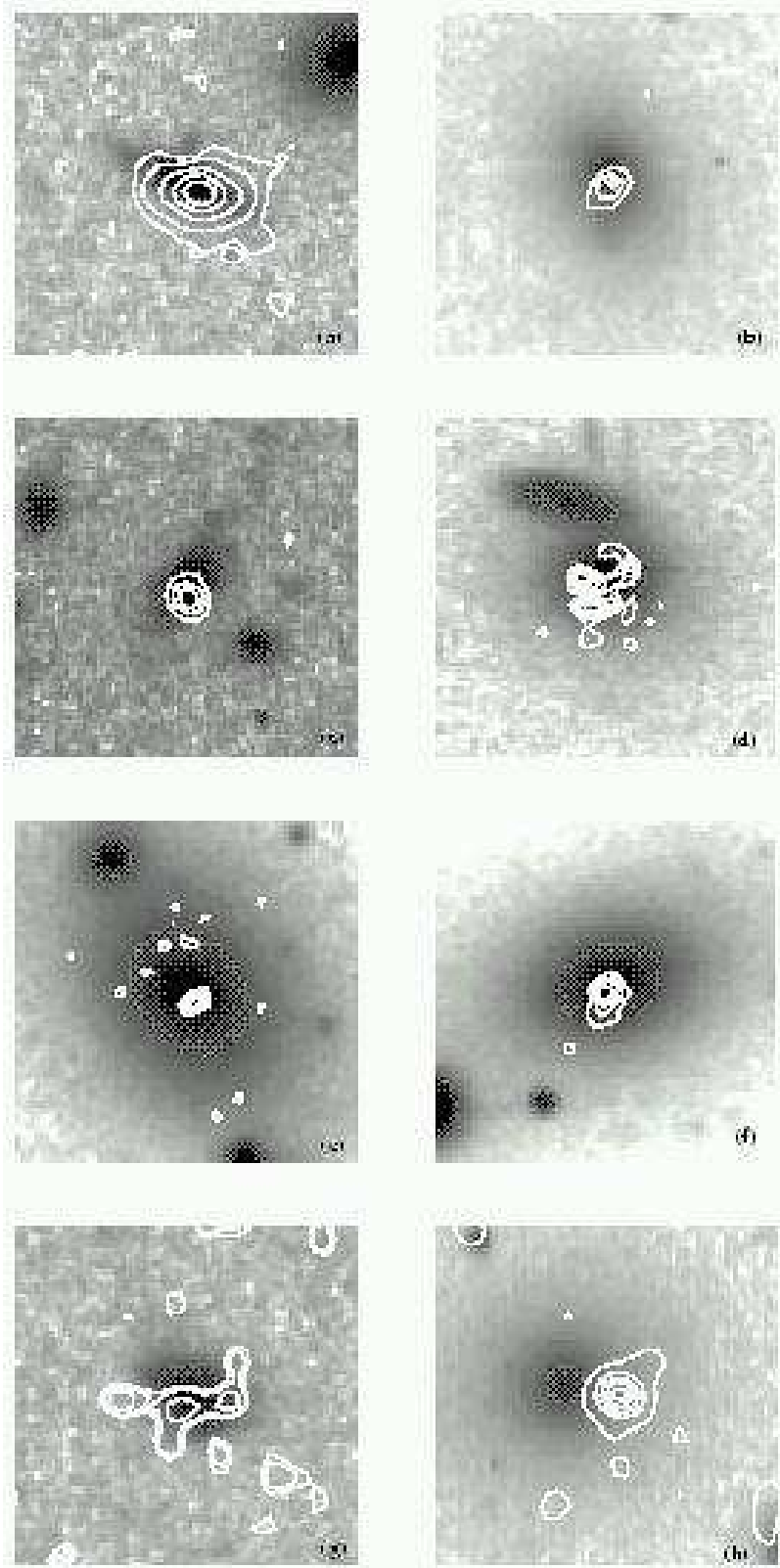}
\caption{
The X-ray contours from the exposure corrected and smoothed ACIS-I
image in the 0.3 -- 10.0 keV band overlaying the optical image from
the KPNO 0.9 meter for the sources in table 1.  The optical field
covers 0.5\arcmin{} on the sky. The image is oriented so that north is
up and east is to the left.  The apparent mismatch between the optical
and X-ray is most likey due to residual coordinate errors.
\label{s1_fig}
}
\end{figure}

The X-ray spectrum for source \#1 is shown in Figure \ref{n1_xray}.
The model fit here is an APEC model for the thermal plasma
and a power law model. The thermal component has a temperature
of 0.15$^{+0.05}_{-0.11}$ keV with a best fit abundance of zero, 
but the upper limit on the abundance is unconstrained. The power
law spectrum has a photon index of 2.98$^{+0.86}_{-1.08}$. For this
fit the absorption is fixed to the Galactic value and the resulting 
$\chi^2$ is 8.8 for 9 degrees of freedom. A pure power law or thermal 
plasma model is a poor fit to the spectrum with a reduced $\chi^2$ of 
1.53 for 11 degrees of freedom. 
The unabsorbed luminosity in the 0.5-10 keV band
is 2.0$\times$10$^{42}$ \ergss. The 2-10 keV luminosity of 
3.7$\times$10$^{41}$ \ergss\, is typical of weak-line radio galaxies
\citep{sambruna99} although the oxygen lines usually seen in these objects 
are not present in the SDSS spectrum (see Figure \ref{fig-spec} and 
Section \ref{sec-opt}). The large fitted photon index has been seen in 
other objects such as NGC~3862 \citep{sun02} and in BL Lac objects 
\citep{cili95,kubo98}. The KPNO image is shown in Figure \ref{s1_fig} 
with the X-ray contours from the smoothed \chandra image overlaying the 
optical.

Sources \#2, \#3, \#7 \& \#8 have too few counts for a spectral fit so
their luminosities are estimated using a power law with the index
fixed to 1.7. We fit the stacked spectra of these sources to test our
assumption that this power law was appropriate and found that combined
spectral fit was consistent with an index of 1.7.
The optical images are shown in Figures \ref{s1_fig}b,
\ref{s1_fig}c, \ref{s1_fig}g and \ref{s1_fig}h. In figure
\ref{s1_fig}h the X-ray peak is offset from the optical center of the
galaxy by 3\farcs 3. We have checked the astrometry by measuring the
position of a known star (N1121133294 from the GSC 2.1) and find it
offset from the X-ray position by $\sim$1\farcs 1.  The average offset
between the point-like X-ray and radio sources is 1\farcs5. Given the 
larger than average offset source
\# 8 may be an ultraluminous X-ray source \citep{mk00}. However, given
the lack of other astrometric references we cannot be certain of this.

The morphology of Source \#4  is more complex and shows signs of possible
extent (Figure \ref{s1_fig}d). The 50\% encircled energy radius for this 
source is 3\farcs 1, larger than the predicted value of $\sim$2\arcsec{} 
at this off-axis angle, and consistent with this being extended thermal 
emission from an elliptical galaxy.  In addition, the spectrum is better fit with a 
thermal plasma than with a power law. The fitted {\sl APEC} model yields a 
temperature of 0.57$^{+0.49}_{-0.32}$ keV and an abundance less than 0.63 
solar which is typical for elliptical galaxies \citep{davis96}.

Source \#5 is fit with fairly large power law index and there is marginal 
evidence in the X-ray spectrum for a soft component like that seen in 
source \#1. However, the net counts for this source are such that the 
inclusion of a soft thermal component is not statistically required and 
the X-ray contours in Figure \ref{s1_fig}e do not show any evidence for 
extended emission. 

Source \#6 has sufficent counts for a fit and the resulting power law is 
typical of active galaxies with a photon index of 1.75$^{+1.90}_{-0.78}$. 
The X-ray contours in Figure \ref{s1_fig}f show a strong point source 
associated with the nucleus of the galaxy.

				\subsection{
Radio
				}\label{sec-radio}

Radio continuum imaging of Abell~2255 was performed using the National
Radio Astronomy Observatory's Very Large Array (VLA).\footnote{The
National Radio Astronomy Observatory is a facility of the National
Science Foundation operated under cooperative agreement by Associated
Universities, Inc.} The observations and reductions are described in
\citet{miller03}. Briefly, the cluster was surveyed at a frequency of
1.4GHz (20cm) using a 25-pointing mosaic. Each pointing had a duration
of $\sim25$ minutes total, and the final map had an rms noise of
$\sim40\mu$Jy beam$^{-1}$ for a 5.9\arcsec{} beam. Adopting a
5$\sigma$ detection limit, over 50 cluster radio galaxies were
identified out to fairly large radii ($\sim40$\arcmin) from the
cluster center.

Abell~2255 has an unusually high number of radio galaxies relative
to other nearby Abell clusters \citep{miller03}. These fall into two
generalized categories: bright elliptical galaxies hosting AGN and
faint spiral and irregular starbursts. The AGN are often spectacular
radio sources, including six sources with extended radio emission.
Four of these lie within the field surveyed by our \chandra 
observations and may be seen in Figure \ref{fig-xrayrad}. Three of
these are ``head-tail'' radio sources in which the host galaxy
lies at one end of the radio emission which extends in a ``tail''
well past the edge of the galaxy, while the fourth is a strong
compact double.

The correspondence of the radio sources and the \chandra sources
is excellent. There are 16 cluster radio galaxies in the general region
covered by the \chandra observations, including all eight of the 
detections. Only three of the four extended radio galaxies are associated 
with \chandra detections; but on closer inspection, the non-detection
of the fourth (J171223+640157) is easily explained as by chance it 
lies between the CCDs of the ACIS-I detector. Hence, eight of 15 cluster 
radio galaxies are also X-ray detections. The typical offset between the 
X-ray and point-like radio positions is 1\farcs 5.

				\subsection{
Optical
				}\label{sec-opt}

Optical spectra for Abell~2255 are available through the SDSS
\citep{stoughton02}. These cover a wavelength range of
3800--9200$\mbox{\AA}$ with a resolving power of $\sim1900$. They are
collected using 640 3\arcsec{} fibers per field, with successive
900-second exposures of a field being obtained until the cumulative
median $(S/N)^2 > 15$ at $g^* = 20.2$ and $i^* = 19.9$ is
achieved. For the region of Abell~2255, this typically amounted to
4500 seconds of exposure time. SDSS spectra were not available for
objects \#5 and \#8. Their spectra, velocity, and classifications 
were obtained from \citet{hill03} and \citet{miller03}.

Figure \ref{fig-spec} depicts the optical spectra of the X-ray
selected cluster galaxies. None of these galaxies show the classical
signs of harboring an AGN: two show the signatures of a starburst,
one shows moderately strong Balmer absorption lines, and the rest have
optical spectra characteristic of a passive galaxy.

				\section{
Discussion 
				}

\begin{figure}
\epsscale{0.9}
%\plotone{LxLBcor.ps}
\plotone{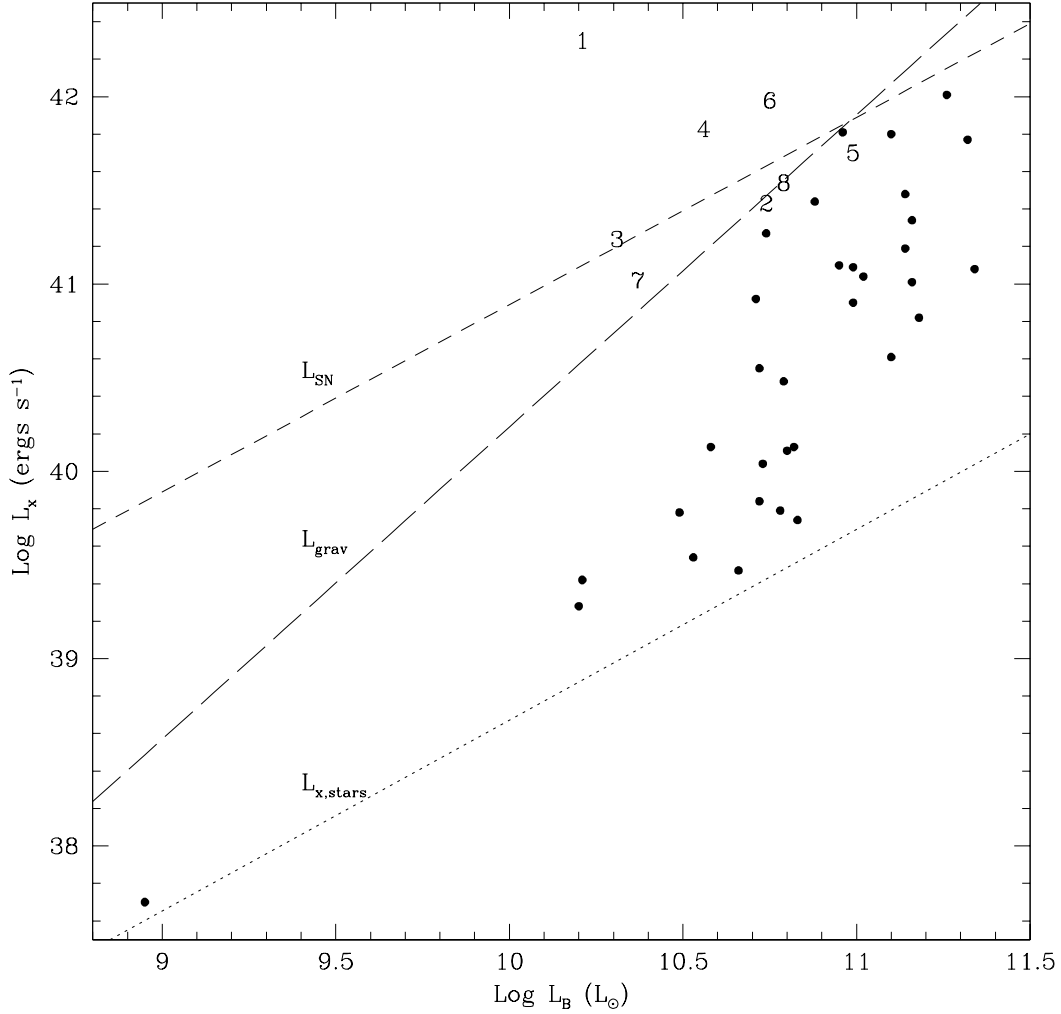}
\caption{
Correlation of X-ray luminosity with optical luminosity, for nearby
normal elliptical galaxies (i.e., non-AGN). The solid points and
expected relationships for X-ray luminosity from stellar emission
(dotted line), gravitational energy release (long dashed line), and
supernovae (dashed line) are taken from \citet{brown98}. Points
corresponding to the eight point X-ray point sources in Abell~2255 are
indicated numerically (see Table \ref{tbl-1} for the
assignments). Their $L_B$ values have been determined from their SDSS
magnitudes using transformations from
\citet{fukugita95}.\label{fig-lxlo}
}
\end{figure}

\begin{figure}
\epsscale{0.9}
%\plotone{CMxray.ps}
\plotone{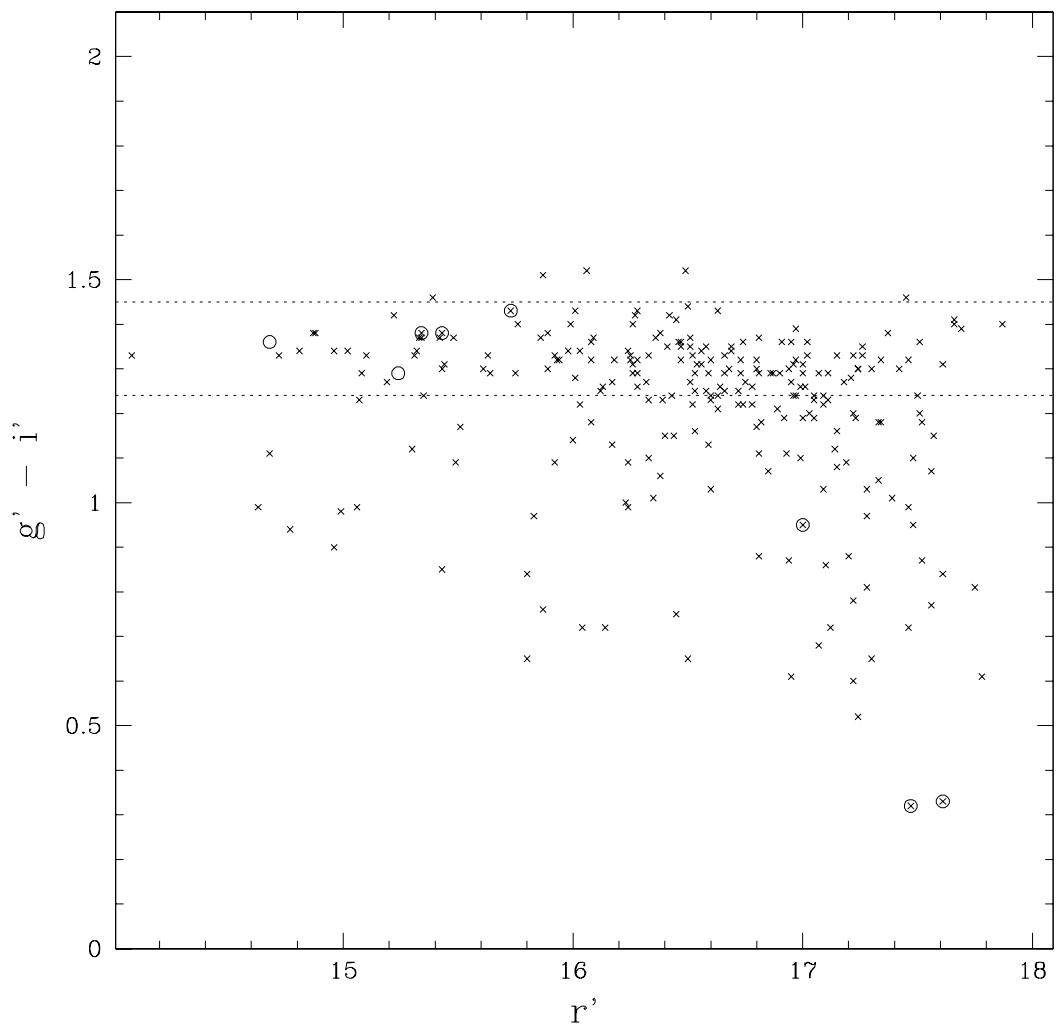}
\caption{
Color magnitude diagram for Abell~2255, based on SDSS data. Only
galaxies with measured SDSS velocities are included. The dotted lines
indicate the expected colors for cluster ellipticals and lenticulars,
based on the transformations of \citet{fukugita95}. The eight galaxies
with associated X-ray point sources are indicated by circles; the
fainter, blue galaxies are the starbursts (\#3 and \#7) and the next 
faintest galaxy is the post-starburst (\#1).\label{fig-cmd}
}
\end{figure}

Searching for AGN using X-ray methods has been very productive.  At
energies above $\sim$2 keV the X-ray sky is dominated by point sources
and most of these are AGN \citep{mushotzky00,barger02}. This fact
prompts us to use the superb angular resolution of \chandra to search
for AGN emission in the eight X-ray detected objects in the central
region of Abell~2255
associated with cluster galaxies. The eight objects cover a wide range
in the \lxlo{} plane with five of them (\#1, \#3, \#4, \#6, and \#7)
having X-ray to optical ratios (\lxlo ) much higher than those of
field elliptical galaxies (see Figure \ref{fig-lxlo}). Objects \#2,
\#5, and \#8 have \lxlo{} consistent with the brighter members of the
\citet{brown98} elliptical galaxy sample. As Figure \ref{fig-cmd} shows, the five
optically brightest X-ray sources seem to be drawn from the normal
population of galaxies in Abell~2255, with $g^* - i^*$ colors similar
to those of cluster elliptical and lenticulars.

The nature of these lower luminosity objects is not certain. We now
know that there can exist ultra-luminous X-ray sources which are not
AGN up to X-ray luminosities $L_X \approx 10^{41}$ \ergss{}
\citep{cp2002}, and it is possible that some of our sources can be
associated with this type of object (e.g. source \# 8). It is also
possible that some of the X-ray emission can be due to hot gas in the
galaxies as seen in the most luminous galaxies in Abell~1367 and the
two massive central galaxies in the Coma cluster
\citep{sun02,vikhlinin01}.  These latter galaxies have X-ray
luminosities of $9.1 \times 10^{40}$ and $7.6 \times 10^{40}$
\ergss{}, similar to our lowest luminosity objects. However, the
uncertainties in the hardness ratios of our sources are not small
enough to confirm that the emission is due to gas in elliptical
galaxies. Despite this, the combined X-ray, optical, and radio data
are of sufficient quality to help constrain the nature of the eight
detected objects.

\begin{figure}
\epsscale{0.9}
%\plotone{Spec.ps}
\plotone{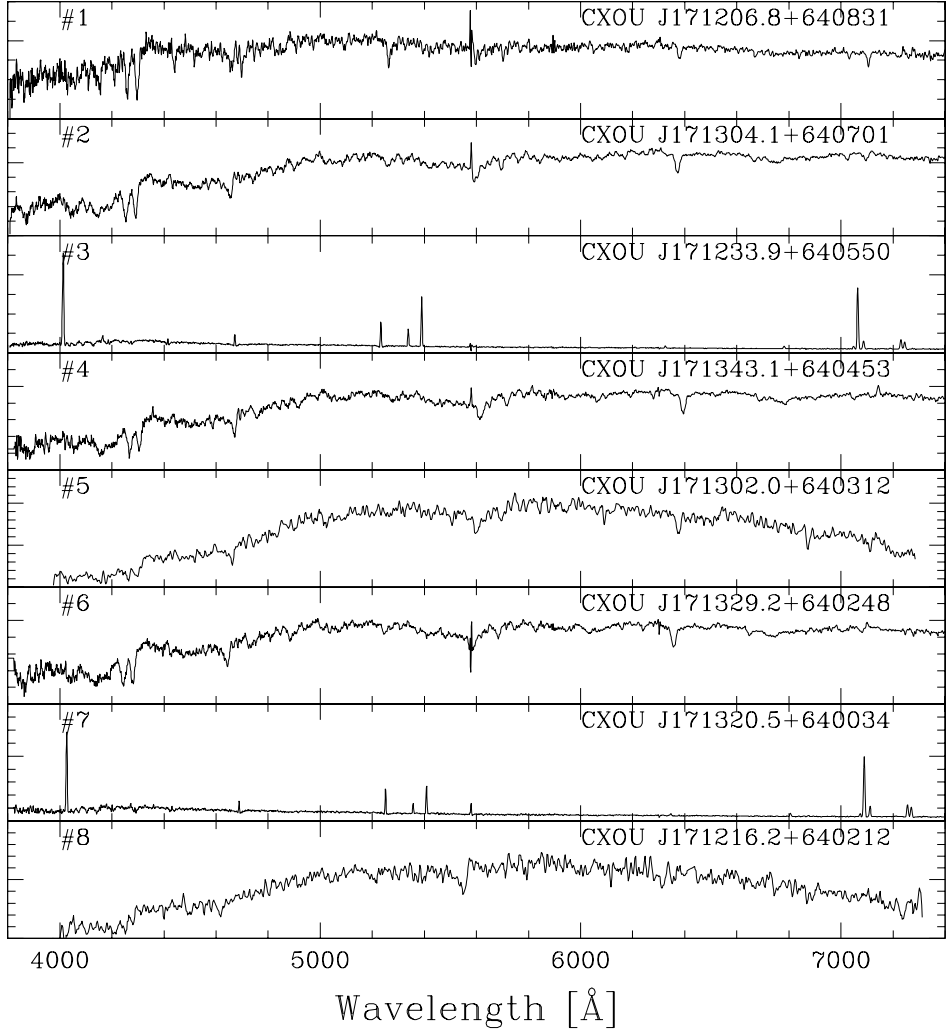}
\caption{
Optical spectra for the X-ray detected cluster galaxies. With the
exception of \#5 and \#8, all are taken from the SDSS. Five of the 
galaxies (\#2, \#4, \#5, \#6, and \#8) are easily distinguished as 
elliptical-like, lacking emission lines and with absorption features 
characteristic of older stellar populations. The spectrum of \#1 is 
that of a post-starburst galaxy, with notable strong Balmer absorption 
and a relatively weak 4000$\mbox{\AA}$ break. The remaining galaxies, 
\#3 and \#7, have classic starburst spectra with strong narrow lines.
The spectra have a resolution of $\sim$1900 and the vertical axis has
been arbitrarily scaled for display. 
\label{fig-spec}
}
\end{figure}

Of the eight objects, we believe that two (\#2, \#6) are
unambiguously associated with AGN and that source \#8 may be an 
AGN. These galaxies have point-like
\chandra emission with $L_X \approx 2.7 \times 10^{41}$ \ergss{},
$L_X \approx 9.5 \times 10^{41}$ \ergss, and $L_X \approx 3.8 
\times 10^{41}$ \ergss{}, respectively, plus hard X-ray
spectra. Each galaxy is a powerful radio source, with \#2 and \#8 being
head-tail sources and \#6 a compact double. The optical spectra
are pure absorption line dominated with no evidence for nuclear
activity (see Figure \ref{fig-spec}).

Source \#1 is rather enigmatic. Its \chandra image is point-like but
at 7\arcmin{} off axis the upper limit on source size is only
2\arcsec. It has a very high \lxlo{} inconsistent with that of normal
elliptical or starburst galaxies, and an optical spectrum indicative
of a post-starburst galaxy (i.e., dominated by Balmer absorption
lines; see Figure \ref{fig-spec}). Its X-ray spectrum is complex,
requiring a thermal plus power law model, but the power law slope is
relatively steep. There are indications from comparison of archival
{\sl ROSAT} data that the source was at least a factor of 3.5 times weaker
when observed in 1993 and 1994 by the {\sl ROSAT} PSPC and HRI. So
the X-ray emission must be dominated by a small
object, indicating that this source is either an AGN, an extremely
luminous binary source (ULX), or perhaps a previously unrecognized
ultra-luminous supernova.

Sources \#3 and \#7 appear to be starburst galaxies. Their X-ray 
emission from the \chandra observation is point-like and they have
hard spectra. Optically, they are both fairly faint galaxies with strong, 
narrow emission lines representative of starbursts. Their radio 
fluxes are consistent with this description, and their X-ray to optical 
ratios are also consistent with starburst activity \citep{ranalli02}. 
The relatively high ratio of hard to soft X-ray flux in source \#3
is unusual for a starburst galaxy, although the lack of a significant 
number of counts limits this statement. Source \#7 is fainter and has
too few counts to even determine a flux ratio. In each case, the implied 
SFRs from the radio and X-ray emission are consistent to within about a 
factor of two. We note that other studies have also identified starburst 
galaxies dominated by very compact X-ray emission 
\citep[e.g., NGC3256][]{ward00}.

The X-ray emission of source \#4 is not easily characterized. It
appears to be slightly extended, with a Gaussian size of
$\sim3$\arcsec{} (6.2$h_{50}^{-1}$ kpc) and a soft X-ray
spectrum. While the galaxy is an interacting system, the X-ray
emission is clearly associated with the southern elliptical
galaxy. Its high \lxlo{} is inconsistent with those of normal
elliptical or starburst galaxies, and its optical spectrum is
dominated by absorption lines. However, the extended nature of the
emission indicates that the source is not dominated by a AGN. We
conclude that this galaxy is similar to the X-ray emission observed
from the cD galaxies in Coma \citep{davis93,vikhlinin01}. It is interesting to
note that without the \chandra evidence, this galaxy would almost
certainly have been classified as an AGN. While many elliptical
galaxies can have X-ray luminosities up to $10^{42}$ \ergss{} produced
by bremmstrahlung from hot gas \citep[e.g.,][]{davis96}, their spectra are very
soft and they are extended on scales of $\sim$10 kpc
\citep{matsushita01}. Observations made with instruments other than
\chandra would fail to resolve the X-ray emission, which when coupled
with its \lxlo{} would result in classification as an AGN. It is
therefore clear that classification based only on \lxlo{} may be
misleading.

Source \#5 appears to be a ``normal'' elliptical galaxy. It is
point-like in the \chandra image, but is relatively weak and
consequently has a poorly determined X-ray spectrum. Optically, its
spectrum, color, and magnitude are typical of cluster elliptical
galaxies. Its X-ray to optical ratio is consistent with those of
nearby non-AGN ellipticals (see Figure \ref{fig-lxlo}), suggesting that
the X-ray emission arises from stars and diffuse hot gas in the
galaxy.

As previously noted, the correspondence between the \chandra detections
and the radio detections is excellent. It is instructive to consider
this in light of the different classifications of the galaxies. 
On the basis of radio and optical properties, \citet{miller03} classify 
eight of the 15 radio galaxies in the ACIS-I field as AGN. Five of these 
are detected in the present \chandra study, with the three non-detections 
being the three weakest radio sources out of the eight (their fluxes 
range from 0.36 mJy to 0.66 mJy; the weakest radio flux of the \chandra
detected AGN is 1.43 mJy). The remaining seven radio sources consist
of six star-forming galaxies and the post-starburst. Interestingly
enough, the two star-forming galaxies detected by the \chandra
observations correspond to the two optically-defined starburst galaxies,
based on EW(\OII)$>40\mbox{\AA}$ \citep{dressler99}. Of the other four 
star-forming galaxies, two have stronger radio emission but are associated 
with larger galaxies whose spectra are more representative of a regular 
star formation history (although the optical spectrum of J171223+640829
has strong similarities to those of starburst galaxies). 

There are several other \chandra papers which report the detection of
point-like X-ray sources in clusters of galaxies: Abell~1367
\citep{sun02}, Abell~2104 \citep{martini02}, Abell~1995
\citep{molnar02}, and RXJ003033.2+261819 and 3C295
\citep{cappi01}. Prior to \chandra, \citet{lazzati98} used {\sl ROSAT}
observations to identify point-like X-ray sources in Abell~194 and
Abell~1367. In all of these papers most of the sources are less
luminous than $3 \times 10^{42}$ \ergss{} (with the exception of one
of the sources in \citet{martini02}) and show no evidence of optical
nuclear activity. However, only Abell~2255 and Abell~1367 are close
enough that the \chandra spatial resolution allows source
classification. In Abell~1367, Sun and Murray report four new AGN and two
new extended sources associated with hot gas, similar to the case in
Abell~2255. Both the luminosity and the classification of the sources
in Abell~1367 are quite similar to that in Abell~2255.

We find that 5\% of the luminous galaxies host low luminosity AGN,
considerably above the values quoted from optical surveys in the
literature \citep[e.g.,][]{dressler99}.  We derive this by combining
the statistics from Abell~2255, Abell~1367 and Abell~2104. To isolate
the AGN the usual method is to select objects with high \lxlo{} and
flat X-ray spectra. But in the case of Abell~2255 it seems clear that
at least one of the objects with high X-ray to optical ratio is a
post-starburst galaxy and one is an elliptical galaxy with $\sim$10x
the normal thermal X-ray flux. 

The implications of the optical spectrum of source \#1 warrant further
discussion. Galaxies with post-starburst optical spectra are tied to
the Butcher-Oemler effect \citep{bo78}, as they are much more common in clusters at
higher redshift than locally \citep[e.g.,][]{dressler99}. Their strong
Balmer absorption implies a burst of star formation that has seemingly
terminated within about the past Gyr. However, the sensitive radio
observations of \citet{smail99} detected a large fraction of the
post-starburst galaxies in the intermediate redshift cluster
Cl~0939+4713. Galaxies with similar radio luminosities are typically
powered by star formation, so the radio data might imply current star
formation whose optical signatures are highly obscured by dust. In
fact, the HST, optical and near-infrared imaging of Smail et al. seemed
to confirm the presence of large amounts of patchy dust
extinction. They thereby concluded that some galaxies with optical
post-starburst spectra might be extreme examples of dust-obscured
starbursts. The X-ray spectrum of the detected post-starburst galaxy
in Abell~2255 is fit using a soft and a hard component, akin to the
spectra of starbursts. Interestingly enough, it has the highest X-ray
luminosity of all the detected sources in Abell~2255 ($2 \times 10^{42}$
\ergss). In the local universe, the only star forming galaxies with
luminosities greater than $10^{42}$ \ergss{} are ultraluminous IR
galaxies \citep{boller99,condon98} with IR luminosities above
$10^{44}$ \ergss{}. Thus, the magnitude of the X-ray emission (and the
power law slope of the hard component) do not fit this picture.
Similarly, the galaxy is detected in radio at a luminosity about an
order of magnitude fainter than the post-starburst radio galaxies in
Cl~0939+4713. This radio emission implies a star formation rate (SFR)
about 50 times less than that derived from its X-ray emission \citep[using
the radio and X-ray SFR relationships found
in][respectively]{yun01,ranalli02}. Thus, while some residual star
formation may be occuring, further explanation for the X-ray emission
is required. It is possible that the high X-ray luminosity (arising
mainly from the hard X-ray component) is caused by a large population of
ULXs related to the recent starburst. However, since the X-ray flux has
changed by a factor of at least 3.5 since the earlier {\sl ROSAT} 
observations this is unlikely. A second possibility is that the starburst 
triggered an AGN in the galaxy. If it is an AGN, its X-ray spectral
characteristics are most consistent with narrow-emission line Seyfert
galaxies \citep{leighly99} which often show both soft and hard
components, or with BL Lac objects. There is one other similar source
to our knowledge, NGC3862 \citep[alias 3C 264,][]{sun02}. This galaxy
has a similar soft X-ray spectra and an optical spectrum lacking
emission lines (but without strong Balmer absorption).

Of the eight galaxies that we have identified as cluster members, four
of these are AGN (three ellipticals and the post-starburst galaxy; note
that source \#5 may also be an AGN) and none would be identified as 
such based on their optical spectrum. From this we deduce that in the 
central 16\arcmin{} of this cluster that the AGN fraction is $\sim$4\% 
(4/106). Adopting a canonical value of $1\%$ based on the optically-derived 
AGN fraction, the probability of there being four AGN among the 106 
cluster galaxies is only $1.8\%$. Despite the low total numbers, this 
is still significant at about 2.4$\sigma$ level. As pointed out by 
\citet{molnar02} this fraction is a lower limit to the number of AGN in 
the cluster since we are not sensitive to AGN with high intrinsic column 
densities.

				\section{
Conclusions
				}

We find that, similar to the \chandra results for Abell~1367 and 
Abell~2104, there
exists a population of point-like or small X-ray sources associated
with the galaxies in Abell~2255. Three of these are clearly AGN, two are
most probably starburst galaxies, one is an X-ray luminous elliptical
dominated by hot gas, and two are of uncertain classification. The
luminosity of the AGN are less than $3 \times 10^{42}$ \ergss{} and
none of the host galaxies show optical evidence for activity. The
most luminous object has a composite X-ray spectrum more reminiscent
of a BL Lac or narrow line Seyfert galaxy and appears to be variable
based on comparison of {\sl ROSAT} and \chandra data. Interestingly,
it is associated with a galaxy whose spectrum is that of a classical
post-starburst. With several clusters now having similar properties we
confirm  the results from Martini et al, that the X-ray
selected AGN fraction in clusters is much higher than optically
selected AGN, similar to the results of the deep \chandra surveys
\citep{barger02}, and these AGN tend to reside in red early-type galaxies. 
This is also consistent with the results of radio
continuum studies.

\acknowledgments
This research made use of the HEASARC, NED, and SkyView databases. NAM
acknowledges the support of a National Research Council Associateship
Award, held at NASA's Goddard Space Flight Center.
We would also like to thank the referee for helpful comments.

\clearpage

\clearpage

\begin{deluxetable}{lrlcrccc}
\footnotesize
\tablecaption{Source Summary\label{tbl-1}}
\tablewidth{0pt}
\tablehead{\colhead{ID} &\colhead{Name}&
\colhead{$\Gamma$\tablenotemark{a}}&\colhead{L$_{\rm x}$} & \colhead{S$_{1.4Ghz}$}& 
\colhead{r$^*$}{}&\colhead{g$^*$-r$^*$} &\colhead{cz (km/s)}}
\startdata
1 &CXOU J171206.8+640831 & 2.98$^{+0.86}_{-1.08}$  & 2.00$\times$10$^{42}$ & 0.457&17.00&0.71&24671 \\
2 &CXOU J171304.1+640701 & 1.7\tablenotemark{b} & 2.68$\times$10$^{41}$ & 65.9\phn\phn&15.34&0.94&24243 \\
3 &CXOU J171233.9+640550 & 1.7\tablenotemark{b} & 1.72$\times$10$^{41}$ & 0.817&17.61&0.11&22838 \\
4 &CXOU J171343.1+640453 & \nodata & 6.69$\times$10$^{41}$ & 2.766 &15.73& 1.00&25355\\
5 &CXOU J171302.0+640312 & 3.71$^{+2.05}_{-1.28}$ & 5.04$\times$10$^{41}$ & 1.430&14.68& 0.97&24623\\
6 &CXOU J171329.2+640248 & 1.75$^{+1.90}_{-0.78}$ & 9.48$\times$10$^{41}$ & 248.3\phn\phn&15.43& 0.90&23470 \\
7 &CXOU J171320.5+640034 & 1.7\tablenotemark{b} & 1.04\phn$\times$10$^{41}$ & 0.535 &17.47& 0.10&23950\\
8 &CXOU J171215.6+640212 & 1.7\tablenotemark{b} & 3.48$\times$10$^{41}$ & 12.6\phn\phn &15.24&0.89&21416\\
\enddata
\tablenotetext{a}{The fit for CXOU J171206.8+640831 includes a thermal plasma.}
\tablenotetext{b}{Parameter is fixed.}
\end{deluxetable}

			\end{document}